\documentclass[a4paper,10pt]{article}%
\usepackage{geometry}
\usepackage{amsmath}
\usepackage{amsfonts}
\usepackage{amssymb}
\usepackage{graphicx}%
\usepackage{indentfirst}
\usepackage[small,bf]{caption}
\providecommand{\U}[1]{\protect\rule{.1in}{.1in}}

\newcommand{\imsuf}{eps}
\newcommand{\p}{\partial}
\newcommand{\mx}{\ensuremath{\mathsf}}
\DeclareMathOperator{\tr}{tr}

\begin{document}

\date{}
\title{\textbf{   $SU(2) \times U(1)$ Yang-Mills theories in $3d$ with Higgs field and Gribov ambiguity}}
\author{\textbf{M.~A.~L.~Capri}$^{a}$\thanks{caprimarcio@gmail.com}\,\,,
\textbf{D.~Dudal}$^{b}$\thanks{david.dudal@ugent.be}\,\,,
\textbf{M.~S.~Guimaraes}$^{a}$\thanks{msguimaraes@uerj.br}\,\,,
\textbf{I.~F.~Justo }$^{a}$\thanks{igorfjusto@gmail.com}\,\,,\\
\textbf{S.~P.~Sorella}$^{a}$\thanks{sorella@uerj.br}\,\,,
\textbf{D.~Vercauteren}$^{a}$\thanks{vercauteren.uerj@gmail.com}\, \thanks{Work supported by
FAPERJ, Funda{\c{c}}{\~{a}}o de Amparo {\`{a}} Pesquisa do Estado do Rio de
Janeiro, under the program \textit{Cientista do Nosso Estado}, E-26/101.578/2010.}\,\,\\[2mm]
{\small \textnormal{$^{a}$  \it Departamento de F\'{\i }sica Te\'{o}rica, Instituto de F\'{\i }sica, UERJ - Universidade do Estado do Rio de Janeiro,}}
 \\ \small \textnormal{\phantom{$^{a}$} \it Rua S\~{a}o Francisco Xavier 524, 20550-013 Maracan\~{a}, Rio de Janeiro, Brasil}\\
	 \small \textnormal{$^{b}$ \it Ghent University, Department of Physics and Astronomy, Krijgslaan 281-S9, 9000 Gent, Belgium}\normalsize}
\maketitle

\begin{abstract}
We study the structure of the gauge propagators  of a $3d$  version of the electroweak interaction in terms of the  Higgs vacuum expectation value $\nu$, of the non-Abelian gauge coupling $g$, and of the Abelian gauge coupling $g'$, when nonperturbative effects related to the non-Abelian gauge fixing are introduced by means of an adapted path integral measure. In the perturbative regime of small non-Abelian coupling $g$ and sufficiently large $\nu$, the well-known standard $Z$ and $W$ propagators are recovered, together  with a massless photon. In general, depending on the relative magnitudes of $g$, $g'$ and $\nu$, we uncover a quite different propagator structure.  In a later stage of research, the results here derived can be used to study the associated phase diagram in more depth.
\end{abstract}

\section{Introduction}

Although the perturbative Higgs mechanism is a fundamental ingredient in the standard model of particle physics,   there are still many unknowns in the study of the strongly coupled dynamics of the Yang--Mills--Higgs system. In a seminal work \cite{Fradkin:1978dv}, Fradkin and Shenker discussed the phase structure of gauge theories in the presence of Higgs fields pointing out the important modifications in the spectrum of the theory as a function of the Higgs field representation and the different regimes of the parameters of the theory, {\it i.e.}~the gauge coupling constant $g$ and the vacuum expectation value of the Higgs field $\nu$. In particular, in the case of Higgs fields in the fundamental representation, Fradkin and Shenker were able to show that the Higgs and the confining phases are smoothly connected. More precisely, from \cite{Fradkin:1978dv}, one learns that these two phases  are separated by a first order line transition which, however, does not extend to the whole phase diagram. Instead, it displays an endpoint. This implies that the Higgs phase can be connected to the confining phase in a continuous way, {\it i.e.}~without crossing the transition line. Even more striking, there exists\footnote{See also refs.\cite{Caudy:2007sf,Greensite:2011zz,Bonati:2009pf}.} \cite{Fradkin:1978dv}  a region in the plane $(\nu,g)$, called the analyticity region, in which the Higgs and confining phases are connected by paths along which the expectation
value of any local correlation function varies analytically, implying the absence of any discontinuity in
the thermodynamical quantities. According to \cite{Fradkin:1978dv}, the spectrum of the theory evolves continuously from
one regime to the other. Said otherwise, the physical states in both regimes are generated by suitable
gauge invariant operators which, in the Higgs phase, give rise to massive bosons, while, in the confining phase, to meson-like states, {\it i.e.}~to bound states of confined excitations.

In a previous work  \cite{Capri:2012cr}  we  investigated the three-dimensional SU(2) Yang--Mills--Higgs theory by taking into account the nonperturbative phenomenon of the Gribov copies \cite{Gribov:1977wm} which are present in the gauge fixing  quantization procedure\footnote{See refs.\cite{Sobreiro:2005ec,Vandersickel:2012tz} for a pedagogical introduction to the Gribov problem.} . In this framework, the dynamics of the model is analyzed through the study of the two-point correlation functions of the gauge fields. A Yukawa-type behavior for the gauge propagator signals that the Higgs phase takes place, while a propagator of Gribov type \cite{Gribov:1977wm,Sobreiro:2005ec,Vandersickel:2012tz}  means that the theory is in a confining phase. We remind here that a propagator of Gribov type does not allow for a particle interpretation, as it exhibits complex  conjugate  poles. As such, it is suitable for the description of a confining phase. In particular, in the case of the fundamental representation for the Higgs field, it was found  \cite{Capri:2012cr}  that there were different regions in parameter space, {\it i.e.}~depending on the values $(g,\nu)$, with the two-point functions displaying a rather different behavior. At weak coupling, corresponding to sufficiently small $g$ and large $\nu$, the Yukawa-type propagators of the perturbative Yang--Mills--Higgs model were recovered.  In some intermediate coupling region, the two-point function still exhibits a pole on the axis of real $p^2$ and with positive residue, meaning there are still physical excitations present in the gauge sector. In this region a second unphysical pole with negative residue was also found. At strong coupling, {\it i.e.}~large $g$ and sufficiently small $\nu$, the effect of the restriction to the Gribov region introduced deep modifications to the two-point functions.  All physical poles disappeared and the two-point function was of the Gribovtype, a feature which can be interpreted as signaling confinement. It is worth to mention here that the poles of the gluon propagator are continuous functions of the parameters $(g,\nu)$. In this sense, the two regions, {\it i.e.}~the Higgs and the confining phases, can be seen as being smoothly connected, in agreement with the general results of \cite{Fradkin:1978dv}.
We find it worthwhile to point out here that the Fradkin-Shenker predictions are settled in a discrete (lattice) version of the gauge-Higgs models, though from the appendix of \cite{Fradkin:1978dv} it is clear that the analyticity argument is built around taking the lattice coupling $\beta\to\infty$ (and thus lattice spacing $a\to0$), \emph{i.e.}~leading to a continuum limit.

The present paper aims at generalizing this analysis to the case of $SU(2) \times U(1)$, which is the gauge group of the electroweak interaction in the Standard Model. In addition to the vacuum expectation value of the Higgs field $\nu$ and the $SU(2)$ gauge coupling constant $g$, we have now a third parameter $g'$ corresponding to the $U(1)$ factor. The three-dimensional case considered here can be used  as a kind of effective description to study the four-dimensional electroweak sector at very high temperatures in the spirit of dimensional reduction\footnote{Note, however, that this dimensional reduction is a delicate process involving a different set of relevant scales and additional (adjoint) Higgs fields related to the temporal gauge fields, ultimately also integrated out if sufficiently massive.} \cite{Kajantie:1995dw}.  As in the previous paper, we study the behavior of the two-point functions of the gauge fields after restricting the path integral to the Gribov region. These two-point functions can then help to uncover information about the elementary excitations present in the theory at different values of the coupling constants.

Our analysis uncovers a very rich behavior of the two-point functions. In all, we distinguish four regions in parameter space, as depicted in \figurename\ \ref{regionsdiag}. At weak non-Abelian coupling $g$, the propagators are all unmodified from the usual perturbative Higgs case, and we recover massive $Z$ and $W$ bosons, and a massless photon. For intermediate values of the non-Abelian coupling constant $g$, the behavior of the two-point functions depends very much on the $U(1)$ coupling $g'$. For sufficiently strong $U(1)$ coupling $g'$, the $W$ bosons are removed from the spectrum, and the two-point function of the off-diagonal fields only has poles away from the axis of real $p^2$. In this region the $Z$ boson and the photon are unaffected by the Gribov horizon and behave as in the perturbative Higgs case. At weak $U(1)$ coupling $g'$, but still intermediate non-Abelian coupling $g$, there is a small region in parameter space where the $W$ bosons are removed as in the previous case, and where the two-point functions of the diagonal $SU(2)$ field and of the Abelian field are modified by the Gribov horizon, resulting in three massive poles on the axis of real $p^2$, two of which have the positive residue necessary to merit the interpretation of a physical excitation. At all other values of the couplings all two-point functions are modified, and we only find one excitation with real mass-squared.  All other poles in the two-point functions are away from the axis of real $p^2$, and can thus, at best, be interpreted as confined.

The paper is organized as follows. Section \ref{sect2} gives a short overview of the Landau gauge fixing in the Yang--Mills--Higgs theory under consideration. In section \ref{sect3} the issue of Gribov copies is recapitulated, and we expand on how to restrict to the Gribov region in our case, giving the gap equations of the Gribov parameters and the two-point functions in the presence of the Gribov restriction. In section \ref{sect4} we make the connection with ref.~\cite{Capri:2012cr} by evaluating the limit of the $U(1)$ coupling constant $g'$ to zero. In section \ref{sect5} the necessity of restricting to the Gribov region is investigated by studying the ghost form factors, and several regions in parameter space are identified. In sections \ref{sect6} and \ref{sect7} the two-point functions are investigated, of the off-diagonal, and of the diagonal and photon fields respectively. Finally, section \ref{concs} concludes the paper.

\section{The action and its gauge fixing} \label{sect2}
\subsection{ Classical considerations}

In this section we will consider the case of $3d$ $SU(2) \times U(1)$ with Higgs field in the fundamental representation.\\
The action of the theory is
\begin{equation}
S=\int d^{3}x\left(\frac{1}{4}F_{\mu \nu }^{a}F_{\mu \nu }^{a}+ \frac{1}{4} B_{\mu\nu} B_{\mu\nu} +
(D_{\mu }^{ij}\Phi ^{j})^{\dagger}( D_{\mu }^{ik}\Phi ^{k})+\frac{\lambda }{2}\left(
\Phi ^{\dagger}\Phi-\nu ^{2}\right) ^{2}\right)  \;,
\label{Sf}
\end{equation}
where the covariant derivative is defined by
\begin{equation}
D_{\mu }^{ij}\Phi^{j} =\partial _{\mu }\Phi^{i} - \frac{ig'}{2}B_{\mu}\Phi^{i} -   ig \frac{(\tau^a)^{ij}}{2}A_{\mu }^{a}\Phi^{j}
\end{equation}
and the vacuum expectation value (\textit{vev}) of the Higgs field is
\begin{equation}
\langle \Phi \rangle  = \left( \begin{array}{ccc}
                                          0  \\
                                          \nu
                                          \end{array} \right)  \;.
\label{vevf}
\end{equation}
The indices $i,j=1,2$ refer to the fundamental representation of $SU(2)$ and $\tau^a, a=1,2,3$, are the Pauli matrices. The coupling constants $g$ and $g'$ refer to the groups $SU(2)$ and $U(1)$, respectively. The field strengths $F^a_{\mu\nu}$ and $B_{\mu\nu}$ are given by
\begin{equation}
F^a_{\mu\nu} = \partial_\mu A^a_\nu -\partial_\nu A^a_\mu + g \varepsilon^{abc} A^b_\mu A^c_\nu \;, \qquad B_{\mu \nu} = \partial_\mu B_\nu -\partial_\nu B_\mu  \;.
\label{fs}
\end{equation}
To obtain the  gauge boson propagators, we consider the quadratic part of the action (\ref{Sf}), given by
\begin{multline}
S_{quad} = \int\!\! d^{3}x \, \frac{1}{2}A_{\mu}^{\alpha}\left[ \left(-\partial_{\mu}\partial_{\mu} + \frac{\nu^{2}}{2}g^{2}\right)\delta_{\mu\nu} + \partial_{\mu}\partial_{\nu} \right]A_{\nu}^{\alpha} + \int\!\! d^{3}x\, \frac{1}{2}B_{\mu}\left[ \left(-\partial_{\mu}\partial_{\mu} + \frac{\nu^{2}}{2}g'^{2}\right)\delta_{\mu\nu} + \partial_{\mu}\partial_{\nu}\right]B_{\nu} \\
+ \int\!\! d^{3}x\, \frac{1}{2}A_{\mu}^{3}\left[ \left(-\partial_{\mu}\partial_{\mu} + \frac{\nu^{2}}{2}g^{2}\right)\delta_{\mu\nu} + \partial_{\mu}\partial_{\nu}\right]A_{\nu}^{3} - \frac{1}{4}\int\!\! d^{3}x\, \nu^{2}g\,g'\,A^{3}_{\mu}B_{\mu} - \frac{1}{4}\int\!\! d^{3}x\, \nu^{2}g\,g'\,B_{\mu}A^{3}_{\mu} \;.
\label{Squad}
\end{multline}
In order to diagonalize expression \eqref{Squad}, we introduce the following fields
\begin{subequations} \begin{gather}
W^+_\mu = \frac{1}{\sqrt{2}} \left( A^1_\mu + iA^2_\mu \right) \;, \qquad W^-_\mu = \frac{1}{\sqrt{2}} \left( A^1_\mu - iA^2_\mu \right)  \;,
\label{ws} \\
Z_\mu =\frac{1}{\sqrt{g^2+g'^2} } \left(  -g A^3_\mu + g' B_\mu \right) \qquad \text{and}\qquad A_\mu =\frac{1}{\sqrt{g^2+g'^2} } \left(  g' A^3_\mu + gB_\mu \right) \;.
\label{za}
\end{gather} \end{subequations}
Let us also give, for further use, the inverse combinations:
\begin{subequations} \begin{gather}
A^1_\mu = \frac{1}{\sqrt{2}} \left( W^+_\mu + W^-_\mu \right) \;, \qquad A^2_\mu = \frac{1}{i\sqrt{2}} \left( W^+_\mu - W^-_\mu \right) \;,
\label{iw} \\
B_\mu =\frac{1}{\sqrt{g^2+g'^2} } \left(  g A_\mu + g' Z_\mu \right) \qquad \text{and}\qquad A^3_\mu =\frac{1}{\sqrt{g^2+g'^2} } \left(  g' A_\mu - gZ_\mu \right) \;.
\label{iza}
\end{gather} \end{subequations}
For the quadratic part of the gauge action, we easily get
\begin{multline}
S_{quad} =  \int d^3 x   \left( \frac{1}{2} (\partial_\mu W^+_\nu - \partial_\nu W^+_\mu)(\partial_\mu W^-_\nu - \partial_\nu W^-_\mu)  + \frac{g^2\nu^2}{2}W^+_\mu W^-_\mu   \right) \\
+ \int d^3x  \left(  \frac{1}{4} (\partial_\mu Z_\nu - \partial_\nu Z_\mu)^2  + \frac{(g^2+g'^2)\nu^2}{4}Z_\mu Z _\mu  +    \frac{1}{4} (\partial_\mu A_\nu - \partial_\nu A_\mu)^2  \right)  \;,
\label{qd}
\end{multline}
from which we can read off the masses of the fields $W^+$, $W^-$, and $Z$:
\begin{equation}
m^2_W = \frac{g^2\nu^2}{2} \;, \qquad m^2_Z =  \frac{(g^2+g'^2)\nu^2}{2}  \;. \label{ms}
\end{equation}

 \subsection{The Landau gauge fixing}
The action \eqref{Sf} has to be supplemented by the gauge fixing term $S_{\text{gf}}$  to allow for a meaningful quantization. To that purpose we adopt the Landau gauge condition,  $\partial_\mu A^a_\mu=0$, $\partial_\mu B_\mu=0 $, which is very helpful in order to deal with the issue of the Gribov copies. Thus
\begin{equation}
S_{\text{gf}} = \int d^3x \; \left( b^a \partial_\mu A^a_\mu + {\bar c}^a \partial_\mu D^{ab}_\mu c^b
+ b\partial_\mu B_\mu + {\bar c}\partial^2 c \right) \;,
\label{cgf}
\end{equation}
where $(b^a, b)$ are the Lagrangian multipliers enforcing the conditions $\partial_\mu A^a_\mu=0$ and $\partial_\mu B_\mu=0 $, and $({\bar c}^a, c^a)$, $({\bar c}, c)$ are the $SU(2)$ and $U(1)$ Faddeev--Popov ghosts respectively. Thus, the total action reads
\begin{equation}
S_{f} = S + S_{\text{gf}}  \;. \label{fact}
\end{equation}
The propagators of the gauge fields are given, in terms of $W^{\pm}_{\mu}$, $Z_{\mu}$ and $A_{\mu}$, as
\begin{subequations} \label{WZA gluon prop} \begin{eqnarray}
\langle W^{+}_{\mu}(p)W^{-}_{\nu}(-p) \rangle &=& \frac{1}{p^{2} + \frac{\nu^{2}}{2}g^{2}}\left(\delta_{\mu\nu} - \frac{p_{\mu}p_{\nu}}{p^{2}}\right) \;, \\
\langle Z_{\mu}(p) Z_{\nu}(-p) \rangle &=& \frac{1}{p^{2} + \frac{\nu^{2}}{2}(g^{2} + g'^{2})} \left(\delta_{\mu\nu} - \frac{p_{\mu}p_{\nu}}{p^{2}}\right)\;, \\
\langle A_{\mu}(p) A_{\nu}(-p) \rangle &=& \frac{1}{p^{2}} \left(\delta_{\mu\nu} - \frac{p_{\mu}p_{\nu}}{p^{2}}\right) \;, \\
\langle A_\mu(p) Z_\nu(-p) \rangle & = & 0 \;.
\end{eqnarray} \end{subequations}
Taking into account the relation between the mass eigenstates and the fields $A^{\alpha}_{\mu}$, $A^{3}_{\mu}$ and $B_{\mu}$, we also have the following propagators,
\begin{subequations} \label{AB gluon prop} \begin{eqnarray}
\langle A^{\alpha}_{\mu}(p) A^{\beta}_{\nu}(-p) \rangle &=& \frac{1}{p^{2} + \frac{\nu^{2}}{2}g^{2}}\left(\delta_{\mu\nu} - \frac{p_{\mu}p_{\nu}}{p^{2}}\right)\delta^{\alpha \beta} \;, \\
\langle A^{3}_{\mu}(p) A^{3}_{\nu}(-p) \rangle &=& \frac{1}{p^{2} + \frac{\nu^{2}}{2}(g^{2} + g'^{2})} \left(\delta_{\mu\nu} - \frac{p_{\mu}p_{\nu}}{p^{2}}\right)\;,  \\
\langle B_{\mu}(p) B_{\nu}(-p) \rangle &=& \frac{1}{p^{2}} \left(\delta_{\mu\nu} - \frac{p_{\mu}p_{\nu}}{p^{2}}\right) \;,  \\
\langle A^{3}_{\mu}(p) B_{\nu}(-p) \rangle &=& \frac{\nu^{2}}{4}(g^{2} + g'^{2}) \frac{g'}{g} \frac{1}{p^{2}\left(p^{2} + \frac{\nu^{2}}{2}(g^{2} + g'^{2}) \right)} \left(\delta_{\mu\nu} - \frac{p_{\mu}p_{\nu}}{p^{2}}\right)\;.
\end{eqnarray} \end{subequations}

\section{The restriction to the Gribov region} \label{sect3}
\label{TRGR}
As established in \cite{Gribov:1977wm}, the gauge fixing procedure is plagued by the existence of Gribov copies,  an observation later on shown to be a general feature of non-Abelian gauge fixing \cite{Singer:1978dk}. In the Landau gauge, it is easy to see that at least the infinitesimally connected gauge copies  can be dealt with by restricting the domain of integration in the path integral to the so called Gribov region $\Omega$ \cite{Gribov:1977wm,Sobreiro:2005ec,Vandersickel:2012tz}, defined as the set of all transverse gauge configurations for which the Faddeev--Popov operator is strictly positive, namely
\begin{equation}
\Omega = \{ A^a_\mu\;, \; \partial_\mu A^a_\mu=0 \;, \; -\partial_\mu D^{ab}_\mu > 0\; \}   \label{gr}
\end{equation}
The region $\Omega$ is known to be convex,  bounded in all directions in field space, to contain the zero gauge field, and to have gauge orbit intersecting it. All these properties ---to our knowledge only proven in the Landau gauge, whence the interest in this gauge--- make $\Omega$ well-chosen as integration domain. We refer to the recent review \cite{Vandersickel:2012tz} for a complete list of references.

The boundary of $\Omega$, where the first vanishing eigenvalue of the Faddeev--Popov operator appears, is called the first Gribov horizon. A way to implement the restriction to the region $\Omega$ has been worked out by Gribov in his original work. It amounts to impose the no-pole condition \cite{Gribov:1977wm,Sobreiro:2005ec,Vandersickel:2012tz} for the connected two-point ghost function $\mathcal{G}^{ab}(k;A) = \langle k | \left(  -\partial D^{ab}(A) \right)^{-1} |k\rangle $, which is nothing but the inverse of the Faddeev--Popov operator $-\partial D^{ab}(A)$. One requires that  $\mathcal{G}^{ab}(k;A)$ has no poles at finite nonvanishing values of $k^2$, so that it stays always positive. That way one ensures that the Gribov horizon is not crossed, {\it i.e.}~one remains inside $\Omega$. The only allowed pole is at $k^2=0$, which has the meaning of approaching the boundary of the region $\Omega$. In a recent work \cite{Capri:2012wx}, the no-pole condition was linked in a more precise fashion to the Zwanziger implementation of the Gribov restriction \cite{Zwanziger:1989mf}, thereby making clear their equivalence.

Following Gribov's procedure \cite{Gribov:1977wm,Sobreiro:2005ec,Vandersickel:2012tz}, for the connected two-point ghost function $\mathcal{G}^{ab}(k;A)$ at first order in the gauge fields,  one finds
\begin{equation}
\mathcal{G}^{ab}(k;A) = \frac{1}{k^{2}}\left[ \delta^{ab} - g^{2}\frac{k_{\mu}k_{\nu}}{k^{2}} \int\!\! \frac{d^{3}p}{(2\pi)^{3}} \,\, \epsilon^{amc}\epsilon^{cnb} \frac{1}{(k-p)^{2}}A^{m}_{\mu}(p)A^{n}_{\nu}(-p)\right]\; ,
\label{ghprop}
\end{equation}
which, owing to
\begin{equation}
\epsilon^{amc}\epsilon^{cnb} = \delta^{an} \delta^{mb}  - \delta^{ab} \delta^{mn} \;, \label{dd}
\end{equation}
can be written as
\begin{equation}
\mathcal{G}^{ab}(k;A) = \left(
  \begin{array}{ll}
   \delta^{\alpha \beta} \mathcal{G}_{off}(k;A) & \,\,\,\,\,\,\,\,0 \\
   \,\,\,\;\;\;\;\;\;0 & \mathcal{G}_{diag}(k;A)
  \end{array}
\right),
\label{gh prop offdiag}
\end{equation}
where
\begin{eqnarray}
\mathcal{G}_{off}(k;A) & = &  \frac{1}{k^{2}} \left[1+ g^{2}\frac{k_{\mu} k_{\nu}}{k^{2}} \int\!\! \frac{d^{3}p}{(2\pi)^{3}} \frac{1}{(k-p)^{2}} \frac{1}{2} \left(A^{\alpha}_{\mu}(p)A^{\alpha}_{\nu}(-p) + 2A^{3}_{\mu}(p)A^{3}_{\nu}(-p)\right)\right]  \nonumber \\
& \simeq&  \frac{1}{k^{2}} \left( \frac{1}{1 - \sigma_{off}(k;A)} \right) \;,
\label{gh off}
\end{eqnarray}
and
\begin{eqnarray}
\mathcal{G}_{diag}(k;A) & = & \frac{1}{k^{2}} \left[1+ g^{2}\frac{k_{\mu} k_{\nu}}{k^{2}} \int\!\! \frac{d^{3}p}{(2\pi)^{3}} \frac{1}{(k-p)^{2}} A^{\alpha}_{\mu}(p)A^{\alpha}_{\nu}(-p)\right] \nonumber \\
& \simeq & \frac{1}{k^{2}} \left( \frac{1}{1 - \sigma_{diag}(k;A)} \right)\;.
\label{gh diag}
\end{eqnarray}
The quantities $\sigma_{off}(k;A)$ and $\sigma_{diag}(k;A)$ turn out to be decreasing functions of $k$ \cite{Gribov:1977wm}. Thus, the no-pole condition is implemented by requiring that
\begin{subequations} \begin{equation}
\sigma_{off}(0;A) < 1  \;,
\label{sigmaoffnopole}
\end{equation}
and
\begin{equation}
\sigma_{diag}(0;A) < 1\;,
\label{sigmadiagnopole}
\end{equation} \end{subequations}
where
\begin{subequations} \begin{equation}
\sigma_{off}(0;A) = \frac{g^{2}}{3} \int\!\! \frac{d^{3}p}{(2\pi)^{3}} \frac{1}{p^{2}} \left( \frac{1}{2} A^{\alpha}_{\mu}(p)A^{\alpha}_{\mu}(-p) + A^{3}_{\mu}(p)A^{3}_{\mu}(-p)\right)  \;,
\label{sigma off}
\end{equation}
and
\begin{equation}
\sigma_{diag}(0;A) = \frac{g^{2}}{3} \int\!\! \frac{d^{3}p}{(2\pi)^{3}} \frac{1}{p^{2}} A^{\alpha}_{\mu}(p)A^{\alpha}_{\mu}(-p)\;.
\label{sigma diag}
\end{equation} \end{subequations}
Expressions  \eqref{sigma off}, \eqref{sigma diag} are obtained by taking the limit $k \rightarrow 0$ of equations \eqref{gh off},  \eqref{gh diag}, and by making use of the property
\begin{eqnarray}
A_{\mu }^{a}(q)A_{\nu }^{a}(-q) &=&\left( \delta _{\mu \nu }-\frac{q_{\mu }q_{\nu }%
}{q^{2}}\right) \omega (A)(q)   \nonumber \\
&\Rightarrow &\omega (A)(q)=\frac{1}{2}A_{\lambda }^{a}(q)A_{\lambda }^{a}(-q)  \label{p1}  \;,
\end{eqnarray}
which follows from the transversality of the gauge field, $q_\mu A^a_\mu(q)=0$. Also, it is useful to remind that, for an arbitrary function $\mathcal{F}(p^2)$, we have
\begin{equation}
\int \frac{d^{3}p}{(2\pi )^{3}}\left( \delta _{\mu \nu }-\frac{%
p_{\mu }p_{\nu }}{p^{2}}\right) \mathcal{F}(p^2)=\mathcal{A}\;\delta _{\mu \nu }  \label{p2} \;,
\end{equation}%
where, upon contracting both sides of eq.\eqref{p2} with $\delta_{\mu\nu}$,
\begin{equation}
\mathcal{A}=\frac{2}{3}\int \frac{d^{3}p}{(2\pi )^{3}}\mathcal{F}(p^2)\;.   \label{p3}
\end{equation}

\subsection{The two gap equations}
In order to implement the restriction to the Gribov region $\Omega$ in the functional integral, we encode the information of the no-pole conditions  \eqref{sigmaoffnopole},\eqref{sigmadiagnopole} into step functions \cite{Gribov:1977wm,Sobreiro:2005ec,Vandersickel:2012tz}:
\begin{equation}
Z=\mathcal{N} \int {[DA DB ]\;\delta (\partial A) \delta({\partial B}) \;\det(\mathcal{\partial D})\;\theta (1-\sigma
_{diag}(A))\; \theta (1-\sigma _{off}(A))\; e^{-S} \;.}
\end{equation}
However,  being mainly interested here in the study of the gauge boson propagators , we will consider the quadratic approximation for the partition function, namely
 \begin{equation}
Z_{quad} = \mathcal{N'} \int [DA DB] \delta (\partial A) \delta({\partial B}) \,\theta(1-\sigma_{off})\, \theta(1-\sigma_{diag})\, e^{-S_{quad}} \;,
\label{Zquad}
\end{equation}
where $S_{quad}$ is the quadratic part of the action (\ref{Sf}). Making use of the integral representation
\begin{equation}
\theta(x) = \int_{-i\infty +\epsilon}^{i\infty +\epsilon} \frac{d\omega}{2\pi i \omega} \; e^{\omega x}   \;, \label{rp}
\end{equation}
we get for the functional integral
\begin{eqnarray}
Z_{quad} &=& \mathcal{N'} \int \frac{d \omega}{2\pi i \omega}\frac{d \beta}{2\pi i \beta} [DA DB] \;\delta (\partial A) \delta({\partial B}) \; \;e^{\omega (1-\sigma_{off})} \, e^{\beta (1-\sigma_{diag})} e^{-S_{quad}} \nonumber \\
&=& \mathcal{N'} \int \frac{d \omega}{2\pi i \omega}\frac{d \beta}{2\pi i \beta} [DA^{\alpha} DA^{3} DB] \; e^{\omega}\,e^{\beta} \exp \left\{-\frac{1}{2} \int\!\! \frac{d^{3}p}{(2\pi)^{3}}\, A^{\alpha}_{\mu}(p) \left[ \left(p^{2} + \frac{\nu^{2}}{2}g^{2} + \right. \right. \right. \nonumber \\
 &{}& \left. \left. + \frac{2g^{2}}{3}\left(\beta +\frac{\omega}{2}\right)\frac{1}{p^{2}}\right) \left(\delta_{\mu\nu} - \frac{p_{\mu}p_{\nu}}{p^{2}}\right)\right]A^{\alpha}_{\nu}(-p) + A^{3}_{\mu}(p) \left[ \left(p^{2} + \frac{\nu^{2}}{2}g^{2} + \frac{2g^{2}}{3}\frac{\omega}{p^{2}}\right) \times \right.\nonumber \\
&{}& \left. \times \left(\delta_{\mu\nu} - \frac{p_{\mu}p_{\nu}}{p^{2}}\right) \right]A^{3}_{\nu}(-p)  + B_{\mu}(p) \left[ \left(p^{2} + \frac{\nu^{2}}{2}g'^{2}\right)\left(\delta_{\mu\nu} - \frac{p_{\mu}p_{\nu}}{p^{2}}\right) \right] B_{\nu}(-p) - \nonumber \\
&{}& \left. - A^{3}_{\mu}(p) \left[\nu^{2}g\,g'\left(\delta_{\mu\nu} - \frac{p_{\mu}p_{\nu}}{p^{2}}\right) \right]B_{\nu}(-p) \right\}\;.
\label{Zquad1}
\end{eqnarray}
It turns out to be convenient to perform the following change of variables
\begin{subequations} \begin{equation}
\beta \to \beta -\frac{\omega}{2}
\end{equation}
and
\begin{equation}
\omega \to \omega\;.
\end{equation} \end{subequations}
Therefore, eq.\eqref{Zquad1} can be rewritten as,
\begin{eqnarray}
Z_{quad} &=& \mathcal{N'} \int \frac{d \omega}{2\pi i}\frac{d \beta}{2\pi i} [DA^{\alpha} DA^{3} DB] \;\delta (\partial A^\alpha) \; \delta (\partial A^3)\; \delta({\partial B}) \;
\; e^{-\ln\left(\beta\omega-\frac{\omega^{2}}{2}\right)} e^{\beta}\,e^{\frac{\omega}{2}}\;  \times \nonumber \\
&{ }& \times \exp \left[-\frac{1}{2} \int \frac{d^{3}p}{(2\pi)^{3}} A^{\alpha}_{\mu}(p)\, Q^{\alpha \beta}_{\mu \nu}\, A^{\beta}_{\nu}(-p)\right] \times
  \exp \left[-\frac{1}{2} \int \frac{d^{3}p}{(2\pi)^{3}} \left(
 \begin{array}{cc}
  A^{3}_{\mu}(p) & B_{\mu}(p)
 \end{array} \right)
\mathcal{P}_{\mu \nu} \left(
\begin{array}{ll}
A^{3}_{\nu}(-p) \\
B_{\nu}(-p)
\end{array} \right) \right]  \nonumber \\
\label{Zquad2}
\end{eqnarray}
where
\begin{subequations} \begin{equation}
Q^{\alpha \beta}_{\mu \nu} = \left[ p^{2} + \frac{\nu^{2} g^{2}}{2} + \frac{2}{3}g^{2}\beta \frac{1}{p^{2}} \right] \delta^{\alpha \beta} \left( \delta_{\mu \nu} - \frac{p_{\mu} p_{\nu}}{p^{2}}\right) \label{Qmunu}
\end{equation}
and
\begin{equation}
\mathcal{P}_{\mu \nu} = \left(
                        \begin{array}{cc}
                         p^{2} + \frac{\nu^{2}}{2}g^{2} + \frac{2\omega}{3}g^{2}\frac{1}{p^{2}} & - \frac{\nu^{2}}{2}g\,g' \\
                         - \frac{\nu^{2}}{2}g\,g' & p^{2} + \frac{\nu^{2}}{2}{g'}^2
                        \end{array} \right) \left( \delta_{\mu \nu} - \frac{p_{\mu} p_{\nu}}{p^{2}} \right)\;.
\label{P mu nu}
\end{equation} \end{subequations}
From expression \eqref{Zquad2} we can easily deduce the two-point correlation functions of the fields $A^\alpha_\mu$, $A^3_\mu$ and $B_\mu$, namely
\begin{subequations} \label{propsaandb} \begin{gather}
\langle  A^{\alpha}_\mu(p) A^{\beta}_\nu(-p) \rangle = \frac{p^2}{p^4 + \frac{\nu^2g^2}{2} p^2 + \frac{2}{3}g^2\beta} \; \delta^{\alpha \beta} \left( \delta_{\mu\nu} - \frac{p_\mu p_\nu}{p^2} \right)  \;,   \label{aalpha}
\\
\langle  A^{3}_\mu(p) A^{3}_\nu(-p) \rangle = \frac{p^2 \left(p^2 +\frac{\nu^2}{2} g'^{2}\right)}{p^6 + \frac{\nu^2}{2} p^4 \left(g^2 +g'^2 \right)  + \frac{\omega g^2}{3} \left( 2p^2 +\nu^2 g'^2 \right)} \;  \left( \delta_{\mu\nu} - \frac{p_\mu p_\nu}{p^2} \right)  \;,   \label{a3a3}
\\
\langle  B_\mu(p) B_\nu(-p) \rangle = \frac{ \left(p^4 +\frac{\nu^2}{2} g^{2} p^2+\frac{2}{3} \omega g^2 \right)}{p^6 + \frac{\nu^2}{2} p^4 \left(g^2 +g'^2 \right)  + \frac{\omega g^2}{3} \left( 2p^2 +\nu^2 g'^2 \right)} \;  \left( \delta_{\mu\nu} - \frac{p_\mu p_\nu}{p^2} \right)  \;,   \label{bb}
\\
\langle  A^3_\mu(p) B_\nu(-p) \rangle = \frac{ \frac{\nu^2}{2} g g'  p^2}{p^6 + \frac{\nu^2}{2} p^4 \left(g^2 +g'^2 \right)  + \frac{\omega g^2}{3} \left( 2p^2 +\nu^2 g'^2 \right)} \;  \left( \delta_{\mu\nu} - \frac{p_\mu p_\nu}{p^2} \right)  \;.   \label{ba3}
\end{gather} \end{subequations}
Moving to the fields $W^{+}_\mu, W^{-}_\mu, Z_\mu, A_\mu$, one obtains
\begin{subequations} \label{propszandgamma} \begin{gather}
\langle  W^{+}_\mu(p) W^{-}_\nu(-p) \rangle = \frac{p^2}{p^4 + \frac{\nu^2g^2}{2} p^2 + \frac{2}{3}g^2\beta} \;  \left( \delta_{\mu\nu} - \frac{p_\mu p_\nu}{p^2} \right)  \;,   \label{ww}
\\
\langle  Z_\mu(p) Z_\nu(-p) \rangle = \frac{\left( p^4 +\frac{2\omega}{3} \frac{g^2 g'^2}{g^2+g'^2}            \right)}{p^6 + \frac{\nu^2}{2} p^4 \left(g^2 +g'^2 \right)  + \frac{\omega g^2}{3} \left( 2p^2 +\nu^2 g'^2 \right)} \;  \left( \delta_{\mu\nu} - \frac{p_\mu p_\nu}{p^2} \right)  \;,   \label{zz}
\\
\langle  A_\mu(p) A_\nu(-p) \rangle = \frac{\left( p^4 +\frac{\nu^2}{2} p^2 (g^2+g'^2) +\frac{2\omega}{3} \frac{g^4}{g^2+g'^2}\right)}{p^6 + \frac{\nu^2}{2} p^4 \left(g^2 +g'^2 \right)  + \frac{\omega g^2}{3} \left( 2p^2 +\nu^2 g'^2 \right)} \;  \left( \delta_{\mu\nu} - \frac{p_\mu p_\nu}{p^2} \right)  \;,   \label{aa}
\\
\langle  A_\mu(p) Z_\nu(-p) \rangle = \frac{\frac{2\omega}{3} \frac{g^3 g'}{g^2+g'^2} }{p^6 + \frac{\nu^2}{2} p^4 \left(g^2 +g'^2 \right)  + \frac{\omega g^2}{3} \left( 2p^2 +\nu^2 g'^2 \right)} \;  \left( \delta_{\mu\nu} - \frac{p_\mu p_\nu}{p^2} \right)  \;.   \label{az}
\end{gather} \end{subequations}
As expected, all propagators get deeply modified in the infrared by the presence of the Gribov parameters $\beta$ and $\omega$. Notice in particular that, due to the parameter $\omega$, a  mixing between the fields $A_\mu$ and $Z_\mu$ arises, eq.\eqref{az}.  The original photon and $Z$ boson as such loose their distinct particle interpretation.  Moreover, it is straightforward to check that in the limit $\beta \rightarrow 0$ and $\omega \rightarrow 0$, the standards propagators are recovered, {\it i.e.}
\begin{subequations} \begin{gather}
\langle  W^{+}_\mu(p) W^{-}_\nu(-p) \rangle {\big |}_{\beta=0} = \frac{1}{p^2 + \frac{\nu^2g^2}{2} } \;  \left( \delta_{\mu\nu} - \frac{p_\mu p_\nu}{p^2} \right)  \;,   \label{ww0}
\\
\langle  Z_\mu(p) Z_\nu(-p) \rangle  {\big |}_{\omega=0} = \frac{1}{p^2 + \frac{\nu^2}{2} \left(g^2 +g'^2 \right)} \;  \left( \delta_{\mu\nu} - \frac{p_\mu p_\nu}{p^2} \right)  \;,   \label{zz0}
\\
\langle  A_\mu(p) A_\nu(-p) \rangle{\big |}_{\omega=0}  = \frac{1}{p^2}   \;  \left( \delta_{\mu\nu} - \frac{p_\mu p_\nu}{p^2} \right)  \;,   \label{aa0}
\\
\langle  A_\mu(p) Z_\nu(-p) \rangle {\big |}_{\omega=0} = 0    \;.   \label{az0}
\end{gather} \end{subequations}
Let us now proceed by deriving the gap equations which will enable us to compute the Gribov parameters $\beta$ and $\omega$ in terms of the parameters $g$, $g'$ and $\nu^2$. To that end we integrate out the fields in expression  \eqref{Zquad2}, obtaining
\begin{equation}
Z_{quad} = \mathcal{N'} \int \frac{d \omega}{2\pi i}\frac{d \beta}{2\pi i}\,e^{-\ln\left(\beta\omega -\frac{\omega^{2}}{2}\right)} e^{\frac{\omega}{2}}\,e^{\beta} \left[\det Q_{\mu \nu}^{ab} \right]^{-1/2}\, \left[ \det \mathcal{P}_{\mu \nu} \right]^{-1/2},  \label{Zquad3}
\end{equation}
with
\begin{subequations} \begin{equation}
\left[\det Q_{\mu \nu}^{ab} \right]^{-1/2} = \exp \left[ -2 \int \frac{d^{3}p}{(2\pi)^{3}} \log \left(p^{2} + \frac{\nu^{2}}{2}g^{2} + \frac{2}{3}g^{2}\beta \frac{1}{p^{2}}\right) \right] \label{calc_det1}
\end{equation}
and
\begin{equation}
\left[ \det \mathcal{P}_{\mu \nu} \right]^{-1/2} = \exp \left[ - \int \frac{d^{3}p}{(2\pi)^{3}} \log \lambda_{+}(p, \omega)\, \lambda_{-}(p, \omega) \right].   \label{calc_det2}
\end{equation} \end{subequations}
where $\lambda_{\pm}$ are the eigenvalues of the $2\times 2$ matrix of eq.\eqref{P mu nu}, {\it i.e.}
\begin{equation}
\lambda_{\pm} = \frac{\left( p^{4} + \frac{\nu^{2}}{4} p^{2}(g^{2} + g'^{2}) + \frac{\omega}{3}g^{2} \right) \pm \sqrt{\left[ \frac{\nu^{2}}{4}(g^{2} + g'^{2})p^{2} + \frac{\omega}{3}g^{2}\right]^{2} - \frac{\omega}{3}\nu^{2}g^{2}\,g'^{2}p^{2}}}{p^{2}}  \;. \label{ev}
\end{equation}
Thus,
\begin{equation}
Z_{quad} = \mathcal{N} \int \frac{d \omega}{2\pi i}\frac{d \beta}{2\pi i} e^{f(\omega, \beta)} \;, \label{Zf eq}
\end{equation}
where
\begin{equation}
f(\omega, \beta) = \frac{\omega}{2} + \beta -2 \int \frac{d^{3}p}{(2\pi)^{3}} \log \left(p^{2} + \frac{\nu^{2}}{2}g^{2} + \frac{2}{3}g^{2}\beta \frac{1}{p^{2}}\right) - \int \frac{d^{3}p}{(2\pi)^{3}} \log \lambda_{+}(p, \omega)\, \lambda_{-}(p, \omega)\;. \label{f eq}
\end{equation}
Following Gribov's framework \cite{Gribov:1977wm,Sobreiro:2005ec,Vandersickel:2012tz}, expression \eqref{Zf eq} is evaluated in a saddle point approximation, {\it i.e.}
\begin{equation}
Z_{quad} \simeq e^{f(\omega^{\ast}, \beta^{\ast})}\;,
\label{Zf eq1}
\end{equation}
where $(\beta^*, \omega^*)$ are determined by the stationarity conditions
\begin{equation}
\frac{\partial f(\omega, \beta)}{\partial \beta}\bigg{|}_{\beta^{\ast}, \omega^{\ast}} = \frac{\partial f(\omega, \beta)}{\partial \omega}\bigg{|}_{\beta^{\ast}, \omega^{\ast}} = 0 \;, \nonumber
\label{saddle condition}
\end{equation}
from which we get the two gap equations: the first one, from the $\omega$ derivative,
\begin{subequations} \begin{multline}
\frac{2}{3} g^{2} \int \frac{d^{4}p}{(2\pi)^{4}} \frac{1}{\left [ p^{4} + \frac{\nu^{2}}{4}(g^{2}+g'^{2})p^{2} + \frac{\omega^{\ast}}{3}g^{2}\right]^{2} - \left[ \frac{\nu^{2}}{4}(g^{2}+g'^{2})p^{2}+\frac{\omega^{\ast}}{3}g^{2}\right]^{2}+\frac{\omega^{\ast}}{3}\nu^{2}g^{2}g'^{2}p^{2}} \times \\
\times \left\{ \left[ 1+ \frac{\frac{\nu^{2}}{4}(g^{2}+g'^{2})p^{2}+\frac{\omega^{\ast}}{3}g^{2}-\frac{\nu^{2}}{2}g'^{2}p^{2}}{\sqrt{\left[ \frac{\nu^{2}}{4}(g^{2}+g'^{2})p^{2}+\frac{\omega^{\ast}}{3}g^{2}\right]^{2} - \frac{\omega^{\ast}}{3}\nu^{2}g^{2}g'^{2}p^{2}}}\right] \right. \times \\
\left[ p^{4}+\frac{\nu^{2}}{4}(g^{2}+g'^{2})p^{2}+\frac{\omega^{\ast}}{3}g^{2} - \sqrt{\left[ \frac{\nu^{2}}{4}(g^{2}+g'^{2})p^{2}+\frac{\omega^{\ast}}{3}g^{2}\right]^{2}-\frac{\omega^{\ast}}{3}\nu^{2}g^{2}g'^{2}p^{2}}\right] + \\
+ \left[ 1-\frac{\frac{\nu^{2}}{4}(g^{2}+g'^{2})p^{2}+\frac{\omega^{\ast}}{3}g^{4}-\frac{\nu^{2}}{2}g'^{2}p^{2}}{\sqrt{\left[ \frac{\nu^{2}}{4}(g^{2}+g'^{2})p^{2}+\frac{\omega^{\ast}}{3}g^{2}\right]^{2} - \frac{\omega^{\ast}}{3}\nu^{2}g^{2}g'^{2}p^{2}}} \right] \times \\
\left. \left[ p^{4}+\frac{\nu^{2}}{4}(g^{2}+g'^{2})p^{2}+\frac{\omega^{\ast}}{3}g^{2} + \sqrt{\left[ \frac{\nu^{2}}{4}(g^{2}+g'^{2})p^{2}+\frac{\omega^{\ast}}{3}g^{2}\right]^{2}-\frac{\omega^{\ast}}{3}\nu^{2}g^{2}g'^{2}p^{2}}\right] \right\} =1\;,
\label{omega gap eq1}
\end{multline}
and the second one, from the $\beta$ derivative,
\begin{equation}
\frac{4}{3}g^{2} \int \frac{d^{3}p}{(2\pi)^{3}} \frac{1}{p^{4}+\frac{\nu^{2}}{2}g^{2}p^{2}+\frac{2}{3}g^{2}\beta^{\ast}} =1\;.
\label{beta gap eq}
\end{equation} \end{subequations}
In particular, after a little algebra, eq.\eqref{omega gap eq1} can be considerably simplified, yielding
\begin{equation}
\frac{4}{3}g^{2}\int\!\! \frac{d^{3}p}{(2\pi)^{3}}\,\frac{p^{2} + \frac{\nu^{2}}{2}g'^{2}}{p^{6} + \frac{\nu^{2}}{2}(g^{2} + g'^{2})p^{4} + \frac{2}{3}\omega^{\ast}g^{2}p^{2} + \frac{\omega^{\ast}}{3}\nu^{2}g^{2}\,g'^{2}} = 1 \;.
\label{omega gap eq}
\end{equation}

\section{Removing the $U(1)$ factor: study of the limit $g' \to 0$} \label{sect4}
It is useful to study the limit $g' \to 0$, in which case we have to recover the case of $SU(2)$ in the fundamental representation, already discussed in  \cite{Capri:2012cr}. This is an important check for the whole procedure. In particular, in the limit $g' \to 0$, we have to recover the gap equation for the $SU(2)$ case in the fundamental representation \cite{Capri:2012cr}. However, before the analysis of the gap equation, it turns out to be interesting to have a look at the propagators. Let us start by considering first the case without Gribov parameters. Taking the limit $g' \to 0$ in equation \eqref{WZA gluon prop}, we immediately obtain
\begin{subequations} \label{WZAg'0} \begin{eqnarray}
\langle W^{+}_{\mu}(p)W^{-}_{\nu}(-p) \rangle &=& \frac{1}{p^{2} + \frac{\nu^{2}}{2}g^{2}}\left(\delta_{\mu\nu} - \frac{p_{\mu}p_{\nu}}{p^{2}}\right) \;, \\
\langle Z_{\mu}(p) Z_{\nu}(-p) \rangle &=& \frac{1}{p^{2} + \frac{\nu^{2}}{2}g^{2}} \left(\delta_{\mu\nu} - \frac{p_{\mu}p_{\nu}}{p^{2}}\right)\;,  \\
\langle A_{\mu}(p) A_{\nu}(-p) \rangle &=& \frac{1}{p^{2}} \left(\delta_{\mu\nu} - \frac{p_{\mu}p_{\nu}}{p^{2}}\right) \;, \\
\langle A_\mu(p) Z_\nu(-p) \rangle & = & 0 \;.
\end{eqnarray} \end{subequations}
We thus see that, taking the limit $g' \to 0$ in the model $SU(2) \times U(1)$, we will get three massive bosons as in the case of $SU(2)$ with a fundamental Higgs, with the addition of a harmless massless photon completely decoupled from the rest of the theory.

Let us now see what happens to  the propagators when the restriction to the Gribov region is taken into account.  From equations \eqref{propszandgamma} we get the following expressions
\begin{subequations} \begin{eqnarray}
\langle  W^{+}_\mu(p) W^{-}_\nu(-p) \rangle &=& \frac{p^2}{p^4 + \frac{\nu^2g^2}{2} p^2 + \frac{2}{3}g^2\beta} \;  \left( \delta_{\mu\nu} - \frac{p_\mu p_\nu}{p^2} \right)  \;,   \label{wwg'0}
\\
\langle  Z_\mu(p) Z_\nu(-p) \rangle &=& \frac{ p^2}{p^4 + \frac{\nu^2 g^2}{2} p^2   + \frac{2 \omega g^2}{3}} \;  \left( \delta_{\mu\nu} - \frac{p_\mu p_\nu}{p^2} \right)  \;,   \label{zzg'0}
\\ \langle  A_\mu(p) A_\nu(-p) \rangle &=& \frac{\left( p^2 +\frac{\nu^2g^2}{2}  \right)}{p^4 + \frac{\nu^2g^2}{2} p^2  + \frac{2 \omega g^2}{3} } \;  \left( \delta_{\mu\nu} - \frac{p_\mu p_\nu}{p^2} \right)  \;,   \label{aag'0}\\
\langle  A_\mu(p) Z_\nu(-p) \rangle &=&  0   \;,   \label{azg'0}
\end{eqnarray} \end{subequations}
which, when rewritten in terms of $A^{\alpha}_\mu, A^3_\mu, B_\mu$, give
\begin{subequations} \begin{eqnarray}
\langle A^{\alpha}_{\mu}(p) A^{\beta}_{\nu}(-p) \rangle &=& \frac{p^{2}}{p^{4} + \frac{\nu^{2}}{2}g^{2}p^{2} + \frac{2}{3}g^{2}\beta}\left(\delta_{\mu \nu} - \frac{p_{\mu}p_{\nu}}{p^{2}}\right)\delta^{\alpha \beta}\;, \label{Aalpha prop2} \\
\langle  A^3_\mu(p) A^3_\nu(-p) \rangle &=& \frac{ p^2}{p^4 + \frac{\nu^2 g^2}{2} p^2   + \frac{2 \omega g^2}{3}} \;  \left( \delta_{\mu\nu} - \frac{p_\mu p_\nu}{p^2} \right)  \;,   \label{a3a3g'0}
\\
\langle  B_\mu(p) B_\nu(-p) \rangle &=& \frac{1}{p^2} \;  \left( \delta_{\mu\nu} - \frac{p_\mu p_\nu}{p^2} \right)  \;,   \label{bbg'0}\\
\langle  A^3_\mu(p) B_\nu(-p) \rangle &=&  0   \;,   \label{a3bg'0}
\end{eqnarray} \end{subequations}
Let us now proceed to take the limit $g' \to 0$ in the two gap equations. For that purpose we notice that,  in the limit $g' \to 0$,    expression  \eqref{Zquad1} can be written as
\begin{eqnarray}
Z_{quad} &=& \mathcal{N'} \int\!\! \frac{d\omega}{2\pi i} \frac{d\beta}{2\pi i} \; [DA^{\alpha} DA^3 DB]\;   \;\delta (\partial A^\alpha) \; \delta (\partial A^3)\; \delta({\partial B}) \; \,e^{-\ln\left(\beta\omega - \frac{\omega^{2}}{2}\right)}\, e^{\frac{\omega}{2} + \beta} \nonumber \\
&{ }&\exp\!\left\{ -\frac{1}{2}\int\!\! \frac{d^{3}p}{(2\pi)^{3}} \, A^{\alpha}_{\mu} \left[ \left(p^{2} + \frac{\nu^{2}}{2}g^{2}  \right.\right. \right.
 \left. \left. + \frac{g^{2}}{2} \beta \frac{1}{p^{2}}\right)\left(\delta_{\mu\nu}-\frac{p_{\mu}p_{\nu}}{p^{2}}\right)\right]A^{\alpha}_{\nu}(-p) + A^{3}_{\mu}(p) \left[ \left( p^{2}+ \frac{\nu^{2}}{2}g^{2} + \right. \right. \nonumber \\
&{}& \left. \left. \left. + \frac{g^{2}\omega}{2}\frac{1}{p^{2}}\right)\left(\delta_{\mu\nu}-\frac{p_{\mu}p_{\nu}}{p^{2}}\right) \right]A^{3}_{\nu}(-p) + B_{\mu}(p)\, p^{2}\left(\delta_{\mu\nu}-\frac{p_{\mu}p_{\nu}}{p^{2}}\right)B_{\nu}(-p) \right\}\;.
\label{func gerador g'nulo}
\end{eqnarray}
Therefore, we can repeat the previous procedure and evaluate the partition function \eqref{func gerador g'nulo} in the saddle point approximation, getting the following gap equations:
\begin{subequations} \begin{equation}
\frac{4}{3}g^{2} \int \frac{d^{3}p}{(2\pi)^{3}} \frac{1}{p^{4}+\frac{\nu^{2}}{2}g^{2}p^{2}+\frac{2}{3}g^{2}\beta^{\ast}} =1\;,
\label{beta gap eq3}
\end{equation}
for the $\beta$ gap equation, and
\begin{equation}
\frac{4}{3}g^{2}\int\!\! \frac{d^{3}p}{(2\pi)^{3}}\,\frac{1}{p^{4} + \frac{\nu^{2}}{2}g^{2} p^{2} + \frac{2}{3}\omega^{\ast}g^{2}} = 1 \;,
\label{omega gap eq3}
\end{equation} \end{subequations}
for the $\omega$ gap equation.

It is apparent thus that the two gap equations \eqref{beta gap eq3} and \eqref{omega gap eq3} coalesce into a unique gap equation, and they give
\begin{equation}
\omega^{\ast} = \beta^{\ast} \;. \label{same}
\end{equation}
As a consequence, the gauge propagators \eqref{Aalpha prop2} and  \eqref{a3a3g'0}  will depend on a single Gribov parameter, say $\omega^{\ast}$. Expressions \eqref{Aalpha prop2}, \eqref{a3a3g'0}and \eqref{omega gap eq3} coincide with those already found in \cite{Capri:2012cr}, meaning that the limit  $g' \to 0$ correctly reproduces  the results obtained in the case of $SU(2)$ with a Higgs field in the fundamental representation.

\section{Evaluation of the ghost form factors} \label{sect5}

Before discussing the gap equations equations \eqref{beta gap eq} and \eqref{omega gap eq}, it is worthwhile to evaluate the vacuum expectation values of the ghost form factors  $\sigma_{off}(0)$ and $\sigma_{diag}(0)$, eqs. \eqref{sigma off} and \eqref{sigma diag}, without taking into account the restriction to the Gribov region, {\it i.e.}~without the presence of the two Gribov parameters $(\beta^{\ast},\omega^{\ast})$. This will enable us to verify if there exist values of the Higgs condensate $\nu$ and of the couplings $(g,g')$ for which both $\langle \sigma_{off}(0) \rangle $ and $\langle \sigma_{diag}(0) \rangle$  already satisfy the no-pole condition
\begin{equation}
\langle \sigma_{off}(0;A) \rangle  < 1  \;, \qquad   \langle \sigma_{diag}(0;A) \rangle < 1\;,
\label{n-pole}
\end{equation}
in which case $\beta^\ast$ and/or $\omega^\ast$ could be immediately set equal to zero.

\begin{figure}\begin{center}
\includegraphics[width=.25\textwidth]{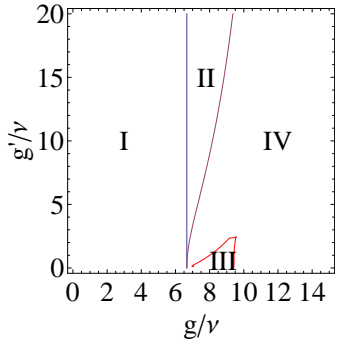}
\caption{There appear to be four regions in phase space. The region I is defined by condition \eqref{secondcond} and is characterized by ordinary Yang--Mills--Higgs behavior (massive $W$ and $Z$ bosons, massless photon). The region II is defined by \eqref{firstcond} while excluding all points of region I --- this region only has electrically neutral excitations, as the $W$ bosons are confined (see Section \ref{sect6}); the massive $Z$ and the massless photon are unmodified from ordinary Yang--Mills--Higgs behavior. Region III has confined $W$ bosons, while both photon and $Z$ particles are massive due to influence from the Gribov horizon; furthermore there is a negative-norm state. In region IV all $SU(2)$ bosons are confined and only a massive photon is left. Mark that the tip of region III is hard to deal with numerically --- the discontinuity shown in the diagram is probably an artefact due to this difficulty.  Details are collected in Section \ref{sect7}. \label{regionsdiag}}
\end{center}\end{figure}

Let us start by considering $\langle \sigma_{off}(0)\rangle$. From eqs.\eqref{sigma off} and \eqref{AB gluon prop} we easily obtain
\begin{eqnarray}
\langle \sigma_{off}(0) \rangle & = &  \frac{2g^{2}}{3}\int\!\!\frac{d^3 p}{(2\pi)^3} \frac{1}{p^{2}}\left(\frac{1}{p^{2} + \frac{\nu^{2}}{2}g^{2}} + \frac{1}{p^{2} + \frac{\nu^{2}}{2}(g^{2}+g'^{2})} \right)  \nonumber \\
 & =& \frac{g^{2}}{3\pi^{2}}\int_{0}^{\infty} \!\!dp\left(\frac{1}{p^{2} + \frac{\nu^{2}}{2}g^{2}} + \frac{1}{p^{2} + \frac{\nu^{2}}{2}(g^{2}+g'^{2})}\right)\;.
\label{sgoff1}
\end{eqnarray}
Analogously, for $\langle \sigma_{diag}(0)\rangle$ one gets
\begin{equation}
\langle \sigma_{diag}(0) \rangle = \frac{4g^{2}}{3}\int\!\!\frac{d^{3}p}{(2\pi)^{3}}\frac{1}{p^{2}}\left(\frac{1}{p^{2} + \frac{\nu^{2}}{2}g^{2}}\right)
= \frac{2g^{2}}{3\pi^{2}}\int_{0}^{\infty}\!\! \frac{dp}{p^{2} + \frac{\nu^{2}}{2}g^{2}}\;.
\label{sgdiag1}
\end{equation}
 As
\begin{eqnarray}
\int_{0}^{\infty} \frac{dp}{p^{2} + m^{2}} &=& \frac{\pi}{2m}\;,
\label{decomp1}
\end{eqnarray}
we found for $\langle \sigma_{off}(0)\rangle$ and $\langle \sigma_{diag}(0)\rangle$
\begin{subequations} \label{rmn1} \begin{eqnarray}
\langle \sigma_{off}(0)\rangle &=& \frac{g}{3\sqrt{2}\pi\nu}(1 + \cos(\theta_{W})) \;, \\
\langle \sigma_{diag}(0)\rangle &=& \frac{2g}{3\sqrt{2}\pi\nu}\;,
\end{eqnarray} \end{subequations} 
where
\begin{equation}
\cos(\theta_{W}) = \frac{g}{\sqrt{g^{2}+g'^{2}}}
\label{thW}
\end{equation}
is the Weinberg angle. For the integration domain of the Yang--Mills field to be the configuration space inside the first Gribov horizon, we need that $\langle \sigma_{off}(0)\rangle$ and $\langle \sigma_{diag}(0)\rangle$  be less than one. Thus,
\begin{subequations} \label{conds} \begin{eqnarray}
(1+\cos(\theta_{W}))\frac{g}{\nu} &<& 3\sqrt{2}\pi \label{firstcond} \\
2\frac{g}{\nu} &<& 3\sqrt{2}\pi \label{secondcond} \;.
\end{eqnarray} \end{subequations}
These two conditions make phase space fall apart in three regions, as depicted in \figurename\ \ref{regionsdiag}.
\begin{itemize}
	\item If $g/\nu<3\pi/\sqrt2$, neither Gribov parameter is necessary to make the integration cut off at the Gribov horizon. In this regime the theory is unmodified from the usual perturbative electroweak theory.
	\item In the intermediate case $3\pi/\sqrt2<g/\nu<3\sqrt2\pi/(1+\cos\theta_W)$ only one of the two Gribov parameters,  $\beta$, is necessary. The off-diagonal ($W$) gauge bosons will see their propagators modified due to the presence of a nonzero $\beta$, while the $Z$ boson and the photon $A$ remain untouched.
	\item In the third phase, when $g/\nu>3\sqrt2\pi/(1+\cos\theta_W)$, both Gribov parameters are needed, and all propagators are influenced by them. The off-diagonal gauge bosons are confined. The behavior of the diagonal gauge bosons depends on the values of the couplings, and the third phase falls apart into two parts, as detailed in section \ref{zandgamma}.
\end{itemize}

\section{The off-diagonal ($W$) gauge bosons} \label{sect6}
Let us first look at the behavior of the off-diagonal bosons under the influence of the Gribov horizon. The propagator \eqref{ww}  only contains the $\beta$ Gribov parameter, meaning that $\omega$ need not be considered here.

As found in the previous section, the parameter $\beta$ is not necessary in the regime $g/\nu<3\pi/\sqrt2$ (region I), due to the ghost form factor $\langle\sigma_{diag}(0)\rangle$ always being smaller than one. In this case, the off-diagonal boson propagator is simply of massive type:
\begin{equation}
  \langle W^{+}_{\mu}(p)W^{-}_{\nu}(-p) \rangle = \frac{1}{p^{2} + \frac{\nu^{2}}{2}g^{2}}\left(\delta_{\mu\nu} - \frac{p_{\mu}p_{\nu}}{p^{2}}\right) \;.
\end{equation}

In the case that $g/\nu>3\pi/\sqrt2$ (regions II, III, and IV), the relevant ghost form factor is not automatically smaller than one anymore, and the Gribov parameter $\beta$ becomes necessary. The value of $\beta^{\ast}$ is determined from the gap equations \eqref{beta gap eq}. After rewriting the integrand in partial fractions, the integral in the equation becomes of standard type, and we readily find the solution
\begin{equation}
  \beta^{\ast} = \frac{3g^2}{32} \left(\frac{g^2}{2\pi^2}-\nu^2\right)^2 \;.
\end{equation}
Mark that, in order to find this result, we had to take the square of both sides of the equation twice. One can easily verify that, in the region $g/\nu>3\pi/\sqrt2$ which concerns us, no spurious solutions were introduced when doing so.

With this value of $\beta^{\ast}$, the off-diagonal propagator can be rewritten as
\begin{multline}
  \langle W_\mu^{+}(p)W_\nu^{-}(-p)\rangle = \frac{\pi/g^3}{\sqrt{\frac{g^2}{4\pi^2}-\nu^2}} \left(\frac{\frac{g^3}{2\pi}\sqrt{\frac{g^2}{4\pi^2}-\nu^2}-\frac i4\nu^2g^2}{p^2+\frac{\nu^2}4g^2+i\frac{g^3}{2\pi}\sqrt{\frac{g^2}{4\pi^2}-\nu^2}} + \frac{\frac{g^3}{2\pi}\sqrt{\frac{g^2}{4\pi^2}-\nu^2}+\frac i4\nu^2g^2}{p^2+\frac{\nu^2}4g^2-i\frac{g^3}{2\pi}\sqrt{\frac{g^2}{4\pi^2}-\nu^2}}\right) \\ \times \left(\delta_{\mu\nu}-\frac{p_\mu p_\nu}{p^2}\right) \;.
\end{multline}
It clearly displays two complex conjugate poles  because $g/\nu>3\pi/\sqrt2$. As such, the off-diagonal propagator  cannot describe a physical excitation of the physical spectrum, being adequate for a confining phase. This means that the off-diagonal components of the gauge field are confined in the region $g/\nu>3\pi/\sqrt2$.

\section{The diagonal $SU(2)$ boson and the photon field} \label{zandgamma} \label{sect7}
The other two gauge bosons --- the $A^3_\mu$ and the $B_\mu$ --- have their propagators given by \eqref{a3a3}, \eqref{bb}, and \eqref{ba3} or equivalently --- the $Z_\mu$ and the $A_\mu$ --- by \eqref{zz}, \eqref{aa} and \eqref{az}. Here, $\omega$ is the only of the two Gribov parameters present.

In the regime $g/\nu<3\sqrt2\pi/(1+\cos\theta_W)$ (regions I and II) this $\omega$ is not necessary to restrict the region of integration to within the first Gribov horizon. Due to this, the propagators are unmodified in comparison to the perturbative case:
\begin{subequations} \label{propsrewrite} \begin{gather}
\langle Z_{\mu}(p) Z_{\nu}(-p) \rangle = \frac{1}{p^{2} + \frac{\nu^{2}}{2}(g^{2} + g'^{2})} \left(\delta_{\mu\nu} - \frac{p_{\mu}p_{\nu}}{p^{2}}\right)\;, \\
\langle A_{\mu}(p) A_{\nu}(-p) \rangle = \frac{1}{p^{2}} \left(\delta_{\mu\nu} - \frac{p_{\mu}p_{\nu}}{p^{2}}\right) \;.
\end{gather} \end{subequations}

In the region $g/\nu>3\sqrt2\pi/(1+\cos\theta_W)$ (regions III and IV) the Gribov parameter $\omega$ does become necessary, and it has to be computed by solving its gap equation, eq. \eqref{omega gap eq}. Due to its complexity it seems impossible to do so analytically. Therefore we turn to numerical methods. Using Mathematica the gap equation can be straighforwardly solved for a list of values of the couplings. Then we determine the values where the propagators have poles. 

The denominators of the propagators are a polynomial which is of third order in $p^2$. There are two cases: there is a small region in parameter space where the polynomial has three real roots, and for all other values of the couplings there are one real and two complex conjugate roots. In \figurename\ \ref{regionsdiag} these zones are labeled III and IV respectively.

Ordinarily, one would like to diagonalize the propagator matrix in order to separate the states present in the theory. In our case, however, doing so requires a nonlocal transformation, and the result will contain square roots containing the momentum of the fields. It seems to be more enlightening to, instead, perform a partial fraction decomposition. If we look at the two-point functions of the $A_3$ and $B$ fields \eqref{a3a3}, \eqref{bb}, and \eqref{ba3}, we can succinctly write those as\footnote{The projector $\delta_{\mu\nu}-\frac{p_\mu p_\nu}{p^2}$ will be ignored in this discussion, as it does not change anything nontrivial here.}
\begin{equation}
	\Delta_{ij} = \frac{f_{ij}(p^2)}{P(p^2)} \;.
\end{equation}
Here, the indices $i,j$ run over $A_3$ and $B$. The functions $f_{ij}(p^2)$ are polynomials of $p^2$ of at most second order, and the function $P(p^2)$ is a third-order polynomial of $p^2$. Furthermore, if we consider the functions $f_{ij}(p^2)$ to be the elements of $2\times2$ matrix, we can see that the determinant of this matrix is nothing but $p^2P(p^2)$.

Let us assume that we know what the roots of $P(p^2)$ are, and call them $-m^2_n$ with $n=1,2,3$. It is then obvious that we can rewrite $P(p^2)$ as $(p^2+m_1^2)(p^2+m_2^2)(p^2+m_3^2)$. We can then perform a decomposition in partial fractions. We will have something of the form
\begin{equation}
	\frac{f_{ij}(p^2)}{P(p^2)} = \sum_{n=1}^3 \frac{\alpha_{ij,n}}{p^2+m_n^2} \;.
\end{equation}
The constants $\alpha_{ij,n}$ can be readily determined the usual way and we get
\begin{equation}
	\frac{f_{ij}(-m_1^2)}{(-m_1^2+m_2^2)(-m_1^2+m_3^2)} = \alpha_{ij,1}
\end{equation}
and analogously for $n=2,3$. In conclusion we find
\begin{multline}
	\Delta_{ij} = \frac{f_{ij}(p^2)}{P(p^2)} = \frac{f_{ij}(-m_1^2)}{(-m_1^2+m_2^2)(-m_1^2+m_3^2)} \frac1{p^2+m_1^2} \\ + \frac{f_{ij}(-m_2^2)}{(m_1^2-m_2^2)(-m_2^2+m_3^2)} \frac1{p^2+m_2^2} + \frac{f_{ij}(-m_3^2)}{(m_1^2-m_3^2)(m_2^2-m_3^2)} \frac1{p^2+m_3^2} \;.
\end{multline}
We can again interpret the constants $f_{ij}(-m_n^2)$ as elements of some $2\times2$ matrices, and we find that the determinants of these matrices are equal to $-m_n^2P(-m_n^2) = 0$, as the $-m_n^2$ are roots of the polynomial $P(p^2)$. Now it is obvious that a $2\times2$ matrix $\mx A$ with zero determinant can always be written in the form $\mx A = v v^T$ with $v$ some $2\times1$ matrix. Furthermore, this vector $v$ has norm $v^Tv = \tr\mx A$. This means that we can write our matrices in the form $\mx A = \tr\mx A \hat v \hat v^T$ where $\hat v$ is now the unit vector parallel to $v$. Therefore, let us write $f_{ij}(-m_n^2) = (f_{11}(-m_n^2)+f_{22}(-m_n^2)) \hat v_i^n \hat v_j^n$, resulting in
\begin{multline} \label{partfracdec}
	\Delta_{ij} = \frac{f_{11}(-m_1^2)+f_{22}(-m_1^2)}{(-m_1^2+m_2^2)(-m_1^2+m_3^2)} \frac1{p^2+m_1^2} \hat v_i^1\hat v_j^1 \\ + \frac{f_{11}(-m_2^2)+f_{22}(-m_2^2)}{(m_1^2-m_2^2)(-m_2^2+m_3^2)} \frac1{p^2+m_2^2} \hat v_i^2\hat v_j^2 + \frac{f_{11}(-m_3^2)+f_{22}(-m_3^2)}{(m_1^2-m_3^2)(m_2^2-m_3^2)} \frac1{p^2+m_3^2} \hat v_i^3\hat v_j^3 \;.
\end{multline}
The vectors $v_i^n$ can be interpreted as linear combinations of the $A_3$ and $B$ fields. Decomposing the two-point functions in this way, we thus find three ``states'' $v_1^n A_3 + v_2^n B$. These states are not orthogonal to each other (which would be impossible for three vectors in two dimensions). The coefficients in front of the Yukawa propagators will be the residues of the poles, and they have to be positive for a pole to correspond to a physical excitation.  The poles can be extracted from the zeros at $p^2_\ast$ of $P(p^2)=p^6+\frac{\nu^2}{2}p^4(g^2+g'^2)+\frac{g^2\omega}{3}(2p^2+\nu^2g'^2)$, viz.
\begin{subequations} \begin{eqnarray}
 p^2_\ast&=& \frac{1}{6}\left\{ (g^{2}+g'^{2})\nu^{2} + \left[ (g^{2}+g'^{2})^{2}\nu^{4} - 8g^{2}\omega\right]\left[ (g^{2}+g'^{2})^{3}\nu^{6} - 12g^{2}(g^{2}-2g^{2})\nu^{2}\omega +  \right. \right. \nonumber \\
&{}&+ \left. 2\sqrt{2}\sqrt{ g^{2}\omega \left( 9g'^{2}(g^{2}+g'^{2})^{3}\nu^{8} - 6g^{2}(g^{4}+20g^{2}g'^{2}-8g'^{4})\nu^{4}\omega + 64g^{4}\omega^{2}  \right)} \right]^{-1/3} + \nonumber \\
&{}&+ \left[ (g^{2}+g'^{2})^{3}\nu^{6} - 12g^{2}(g^{2}-2g^{2})\nu^{2}\omega + 2\sqrt{2}\left( g^{2}\omega \left( 9g'^{2}(g^{2}+g'^{2})^{3}\nu^{8} - \right. \right. \right. \nonumber \\
&{}&- \left. \left. \left. \left. 6g^{2}(g^{4}+20g^{2}g'^{2}-8g'^{4})\nu^{4}\omega + 64g^{4}\omega^{2}  \right)\right)^{1/2}  \right]^{1/3} \right\}
\label{m1}
\end{eqnarray}
\begin{eqnarray}
 p^2_\ast&=& \frac{1}{6}\left\{  (g^{2}+g'^{2})\nu^{2} - \frac{1}{2} \left[ (g^{2}+g'^{2})^{2}\nu^{4} - 8g^{2}\omega\right]\left[ (g^{2}+g'^{2})^{3}\nu^{6} - 12g^{2}(g^{2}-2g^{2})\nu^{2}\omega +   \right. \right. \nonumber \\
&{}&+ \left. 2\sqrt{2}\sqrt{ g^{2}\omega \left( 9g'^{2}(g^{2}+g'^{2})^{3}\nu^{8} - 6g^{2}(g^{4}+20g^{2}g'^{2}-8g'^{4})\nu^{4}\omega + 64g^{4}\omega^{2}  \right)} \right]^{-1/3} - \nonumber \\
&{}&- \frac{1}{2}\left[ (g^{2}+g'^{2})^{3}\nu^{6} - 12g^{2}(g^{2}-2g^{2})\nu^{2}\omega + 2\sqrt{2}\left( g^{2}\omega \left( 9g'^{2}(g^{2}+g'^{2})^{3}\nu^{8} - \right. \right. \right. \nonumber \\
&{}&- \left. \left. \left. \left. 6g^{2}(g^{4}+20g^{2}g'^{2}-8g'^{4})\nu^{4}\omega + 64g^{4}\omega^{2}  \right)\right)^{1/2}  \right]^{1/3} \right. \nonumber \\
&{}&+ i\frac{\sqrt{3}}{2}\left[ (g^{2}+g'^{2})^{2}\nu^{4} - 8g^{2}\omega\right]\left[ (g^{2}+g'^{2})^{3}\nu^{6} - 12g^{2}(g^{2}-2g^{2})\nu^{2}\omega +  \right. \nonumber \\
&{}&+ \left. 2\sqrt{2}\sqrt{ g^{2}\omega \left( 9g'^{2}(g^{2}+g'^{2})^{3}\nu^{8} - 6g^{2}(g^{4}+20g^{2}g'^{2}-8g'^{4})\nu^{4}\omega + 64g^{4}\omega^{2}  \right)} \right]^{-1/3} + \nonumber \\
&{}&+ i\frac{\sqrt{3}}{2}\left[ (g^{2}+g'^{2})^{3}\nu^{6} - 12g^{2}(g^{2}-2g^{2})\nu^{2}\omega + 2\sqrt{2}\left( g^{2}\omega \left( 9g'^{2}(g^{2}+g'^{2})^{3}\nu^{8} - \right. \right. \right. \nonumber \\
&{}&- \left. \left. \left. \left. 6g^{2}(g^{4}+20g^{2}g'^{2}-8g'^{4})\nu^{4}\omega + 64g^{4}\omega^{2}  \right)\right)^{1/2}  \right]^{1/3} \right\}\;,
\label{m2}
\end{eqnarray}
and
\begin{eqnarray}
 p^2_\ast &=& \frac{1}{6}\left\{  (g^{2}+g'^{2})\nu^{2} - \frac{1}{2} \left[ (g^{2}+g'^{2})^{2}\nu^{4} - 8g^{2}\omega\right]\left[ (g^{2}+g'^{2})^{3}\nu^{6} - 12g^{2}(g^{2}-2g^{2})\nu^{2}\omega +   \right. \right. \nonumber \\
&{}&+ \left. 2\sqrt{2}\sqrt{ g^{2}\omega \left( 9g'^{2}(g^{2}+g'^{2})^{3}\nu^{8} - 6g^{2}(g^{4}+20g^{2}g'^{2}-8g'^{4})\nu^{4}\omega + 64g^{4}\omega^{2}  \right)} \right]^{-1/3} - \nonumber \\
&{}&- \frac{1}{2}\left[ (g^{2}+g'^{2})^{3}\nu^{6} - 12g^{2}(g^{2}-2g^{2})\nu^{2}\omega + 2\sqrt{2}\left( g^{2}\omega \left( 9g'^{2}(g^{2}+g'^{2})^{3}\nu^{8} - \right. \right. \right. \nonumber \\
&{}&- \left. \left. \left. \left. 6g^{2}(g^{4}+20g^{2}g'^{2}-8g'^{4})\nu^{4}\omega + 64g^{4}\omega^{2}  \right)\right)^{1/2}  \right]^{1/3} \right. \nonumber \\
&{}&- i\frac{\sqrt{3}}{2}\left[ (g^{2}+g'^{2})^{2}\nu^{4} - 8g^{2}\omega\right]\left[ (g^{2}+g'^{2})^{3}\nu^{6} - 12g^{2}(g^{2}-2g^{2})\nu^{2}\omega +  \right. \nonumber \\
&{}&+ \left. 2\sqrt{2}\sqrt{ g^{2}\omega \left( 9g'^{2}(g^{2}+g'^{2})^{3}\nu^{8} - 6g^{2}(g^{4}+20g^{2}g'^{2}-8g'^{4})\nu^{4}\omega + 64g^{4}\omega^{2}  \right)} \right]^{-1/3} + \nonumber \\
&{}&- i\frac{\sqrt{3}}{2}\left[ (g^{2}+g'^{2})^{3}\nu^{6} - 12g^{2}(g^{2}-2g^{2})\nu^{2}\omega + 2\sqrt{2}\left( g^{2}\omega \left( 9g'^{2}(g^{2}+g'^{2})^{3}\nu^{8} - \right. \right. \right. \nonumber \\
&{}&- \left. \left. \left. \left. 6g^{2}(g^{4}+20g^{2}g'^{2}-8g'^{4})\nu^{4}\omega + 64g^{4}\omega^{2}  \right)\right)^{1/2}  \right]^{1/3} \right\}\;,
\label{m3}
\end{eqnarray} \end{subequations}

It is obviously not possible to disentangle the three (vector-like) degrees of freedom corresponding to these different masses using only two fields. Nonetheless, it is possible to derive a diagonalization of the $1PI$ propagator matrix. After the saddle point approximation and using the $A_\mu$ and $Z_\mu$ field variables, it is not difficult to see that the tree level $(A_\mu,Z_\mu)$ sector of the action arising from eq.~(\ref{Zquad1}) can be reformulated as
\begin{multline}\label{dd1}
  \int d^4p \left(\frac{1}{2}Z_\mu(p)(p^2+g^2\nu^2)Z_\mu(-p)+\frac{1}{2}A_\mu(p)p^2A_\mu(-p)+i\frac{1}{2}\sqrt{\frac{2}{3}\omega_\ast}\left((g'A_\mu(p)-gZ_\mu(p))V_\mu(-p)+(p\leftrightarrow-p)\right)\right. \\ \left.+ \frac{1}{2}V_\mu(p)p^2V_\mu(-p)\right)
\end{multline}
while working immediately on-shell, viz.~using $\p_\mu A_\mu=\p_\mu Z_\mu=0$. The equivalence with the original action can be straightforwardly established by integrating over the $V_\mu$ field. Here, we introduced the latter auxiliary field by hand, but it can be shown in general --- at least for the pure Yang--Mills case; for the current Yang--Mills--Higgs generalization this deserved further investigation at a later stage --- that the all-order no pole condition can be brought in local form by introducing a suitable set of boson and fermion auxiliary fields\footnote{These ghost fields are necessary to eliminate the determinant when integrating over the extra fields.}, see e.g.~\cite{Vandersickel:2012tz,Zwanziger:1989mf,Baulieu:2009ha,Capri:2012wx}.

Having now 3 fields at our disposal with still 3 masses, there is better hope to diagonalize the previous action. First of all, the special limits $g'\to0$ and/or $\omega\to0$ are simply clear at the level of the action (\ref{dd1}). Secondly, the $1PI$ propagator matrix of (\ref{dd1}) only displays a $p^2$-dependence on the diagonal, each time of the form\footnote{In the formulation without the $V_\mu$ field this is not the case. One of the consequences is the appearance of the aforementioned momentum dependent square roots when a diagonalization in terms of the two fields $A_\mu$ and $Z_\mu$ is attempted.} $p^2+\ldots$. As such, the 3 eigenvalues will be of the form $p^2+m_i^2$. Upon using the associated eigenvectors, the action (\ref{dd1}) can then be simply diagonalized to
\begin{equation}\label{dd2}
  \int d^4p \left(\frac{1}{2}\lambda_\mu(p)(p^2+m_1^2)\lambda_\mu(-p)+\frac{1}{2}\eta_\mu(p)(p^2+m_2^2)\eta_\mu(-p)+\frac{1}{2}\kappa_\mu(p)(p^2+m_3^2)\kappa_\mu(-p)\right)\,,
\end{equation}
where the $\lambda_\mu,\eta_\mu,\kappa_\mu$ are the ``generalized $i$-particles'' of the current model, adopting the language of \cite{Baulieu:2009ha}. They are related to the original fields $A_\mu$, $Z_\mu$ and $V_\mu$ by momentum-independent linear transformations. The quadratic form appearing in \eqref{dd2} displays a standard propagator structure, with the possibility that two of the mass poles can be complex conjugate.

\subsection{Three real roots (region III)}
\begin{figure}\begin{center}
\includegraphics[width=.5\textwidth]{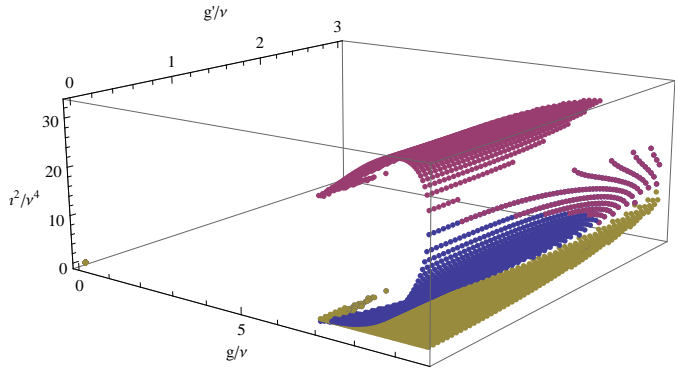}
\caption{The mass-squareds of the massive excitations found in the region where there are three massive poles (region III). \label{threems}}
\end{center}\end{figure}
Region III is defined by the polynomial in the denominators of \eqref{a3a3}, \eqref{bb}, and \eqref{ba3} having three real roots. This region is sketched in \figurename\ \ref{regionsdiag}. (Mark that the tip of the region is distorted due to the difficulty in accessing this part numerically.) The square of the masses corresponding to these three roots are plotted in \figurename\ \ref{threems}.

We computed the residues of these poles, the expression whereof can be read off in the partial fraction decomposition \eqref{partfracdec}. Only the two of the three roots we identified have a positive residue and can correspond to physical states, being the one with highest and the one with lowest mass squared. The third of the roots, the one of intermediate value, has negative residue and thus belongs to some negative-norm state, which cannot be physical.

All three states have nonzero mass for nonzero values of the electromagnetic coupling $g'$, with the lightest of the states becoming massless in the limit $g'\to0$. In this limit we recover the behavior found in this regime in the pure $SU(2)$ case \cite{Capri:2012cr} (the $Z$-boson field having one physical and one negative-norm pole in the propagator) with a massless fermion decoupled from the non-Abelian sector.

\subsection{One real root (region IV)}
\begin{figure}\begin{center}
\includegraphics[width=.5\textwidth]{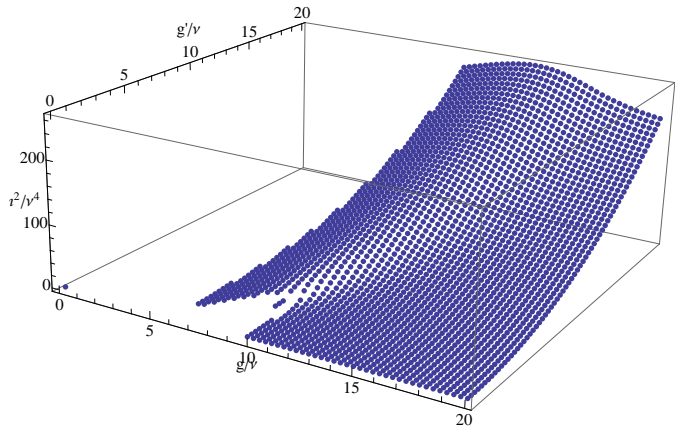}
\caption{The mass-squared of the one physical massive excitation found in region IV. \label{realmass}}
\end{center}\end{figure}

\begin{figure}\begin{center}
\includegraphics[width=.5\textwidth]{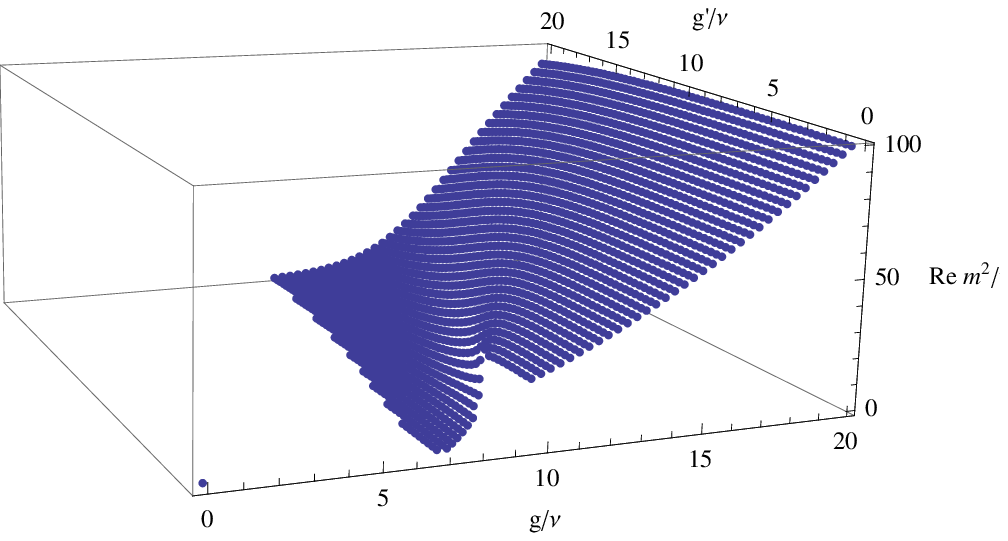}\includegraphics[width=.5\textwidth]{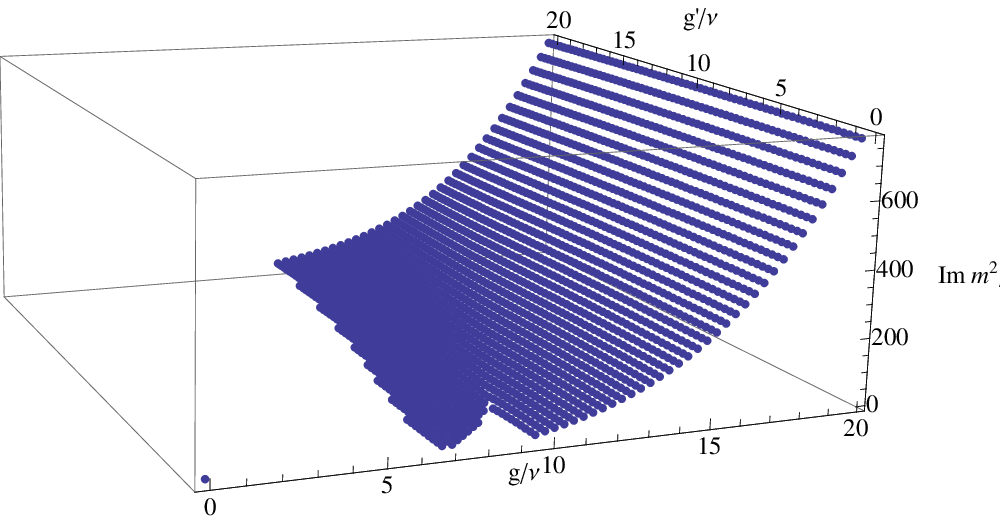}
\caption{The real (left) and imaginary (right) parts of the mass-squared of the other two, complex conjugate, poles. \label{ccmass}}
\end{center}\end{figure}

In the remaining part of parameter space, there is only one state with real mass-squared. The two other roots of the polynomial in the denominators of \eqref{a3a3}, \eqref{bb}, and \eqref{ba3} have nonzero imaginary part and are complex conjugate to each other. The square of the masses corresponding to these roots are plotted in \figurename\ \ref{realmass} for the real root, and in \figurename\ \ref{ccmass} (real and imaginary part) for the complex conjugate roots. In order to determine whether the pole coming from the real root corresponds to a physical particle excitation, we computed its residue, which can be read off in the partial fraction decomposition \eqref{partfracdec}. 

\begin{figure} \begin{center}
\includegraphics[width=.5\textwidth]{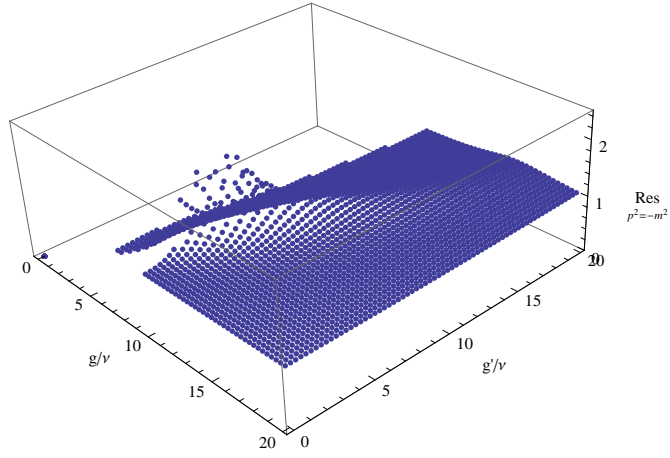}
\caption{The residue of the pole of the photon propagator which is depicted in \figurename\ \ref{realmass}. It turns out to be positive for all values of the couplings within the region IV. \label{resrealmass}}
\end{center} \end{figure}

The result is plotted in \figurename\ \ref{resrealmass}. It turns out the residue is always positive, meaning that this excitation has positive norm and can thus be interpreted as a physical, massive particle. The poles coming from the complex roots cannot, of course, correspond to such physical excitations.

In the limit $g'\to0$ we once more recover the corresponding results already found in the pure $SU(2)$ case \cite{Capri:2012cr} (two complex conjugate poles in the propagator of the non-Abelian boson field) plus a massless photon not influenced by the non-Abelian sector.

\section{Conclusion} \label{concs}
In this work the dynamics of $3d$ Yang--Mills theories in presence of Higgs fields has been investigated from the point of view of the Gribov issue, \emph{i.e.}~by taking into account the existence of the Gribov copies. As a consequence of the restriction of the domain of integration in the functional integral to the Gribov region, the propagator of the gluon field gets considerably modified by the presence of the nonperturbative Gribov parameters (that is the proximity of the Gribov horizon) as well as of the vacuum expectation value of the Higgs field. Looking at the structure of the propagators, we are able to distinguish different regions in parameter space for the physical spectrum of the theory. These regions are depicted in \figurename\ \ref{regionsdiag}.

Region I is the region of weak (non-Abelian) coupling, where the propagators are unmodified from their perturbative Yang--Mills--Higgs behavior. Region II is characterized by the $W$ boson being removed from the spectrum, while the $Z$ boson and the photon are untouched by the restriction to the Gribov region. In this region, the two-point function of the off-diagonal fields is of Gribov type and it does not have any real poles. In regions III and IV the off-diagonal two-point functions are still of Gribov type, but now the two-point functions of the diagonal fields and of the Abelian field are also affected. In region III the combined two-point function of those last two fields has three massive poles, of which two have positive residue --- meaning it may correspond to a physical excitation --- and one has a negative residue --- meaning it cannot possibly be physical. In region IV only one real massive pole is left, and all other poles have nonzero imaginary part.

In this paper, we restricted ourselves to the study of the various gauge boson propagators using the generalization of the semi-classical Gribov no-pole analysis. In subsequent work, we plan to generalize to $4d$ and to construct the local version of the restriction \emph{\`{a} la} Zwanziger, see e.g.~\cite{Vandersickel:2012tz}, and to investigate, using the propagator set constructed here, the genuine phase diagram of gauge-Higgs systems using an appropriate (quasi)order parameter like the Polyakov loop \cite{Greensite:2011zz}. That Gribov quantization can shed light on the phase diagram and on thermodynamics was recently pointed out in \cite{Fukushima:2012qa,Fukushima:2013xsa}; the fact that we revealed a rich structure of the input propagators here in terms of the couplings and Higgs vacuum expectation value suggests an equally rich phase diagram might eventually emerge. We hope to came back to this in future work.

\section*{Acknowledgments}
The Conselho Nacional de Desenvolvimento Cient\'{\i}fico e
Tecnol\'{o}gico (CNPq-Brazil), the Faperj, Funda{\c{c}}{\~{a}}o de
Amparo {\`{a}} Pesquisa do Estado do Rio de Janeiro, the Latin
American Center for Physics (CLAF), the SR2-UERJ,  the
Coordena{\c{c}}{\~{a}}o de Aperfei{\c{c}}oamento de Pessoal de
N{\'{\i}}vel Superior (CAPES)  are gratefully acknowledged. D.~D.~is supported by the Research-Foundation Flanders.

\end{document}